\documentclass[epj,final]{svjour}
\usepackage{graphics,psfig}

 \newcommand\beq{\begin{equation}}
 \newcommand\eeq{\end{equation}}
 \newcommand\beqn{\begin{eqnarray}}
 \newcommand\eeqn{\end{eqnarray}}

 \def\inf{\int_{-\infty}^{\infty}}
 
\begin{document}

\title{Nuclear Shadowing in DIS at Moderately Small $x_B$ }

\author{
J.~Raufeisen\inst{1}\fnmsep
\thanks{\email{J.Raufeisen@mpi-hd.mpg.de}
}
\and
A.V.~Tarasov\inst{1,2}\fnmsep
\thanks{
\email{avt@dxnhd1.mpi-hd.mpg.de}
}
\and
O.O.~Voskresenskaya\inst{2}\fnmsep
\thanks{
\email{voskr@cv.jinr.dubna.su}
}
}

\institute{Institut f\"ur Theoretische Physik,
	Phi\-lo\-so\-phen\-weg 19,
	69120 Heidelberg, Germany 
	\and
	Joint
Institute
for
Nuclear
Research,
Dubna,
141980
Mos\-cow
Region,
Russia.
}

\date{Draft: 9. March 1999}

\abstract{
In the rest frame of the nucleus, shadowing is due to hadronic fluctuations of
the incoming virtual photon, which interact with the nucleons.
We expand these fluctuations in a basis of eigenstates of the 
interaction and take
only the $q\bar q$ component of the hadronic structure of the photon into
account. We use a representation in which the $q\bar q$-pair has a definite
transverse size. 
Starting from the Dirac equation, we develop a path integral approach that
allows to sum all multiple scattering terms and accounts for fluctuations of the
transverse size of the pair, as well as for the finite lifetime of the hadronic
state.  
First numerical results 
show that 
higher order scattering terms have a strong influence on the total 
cross section $\sigma^{\gamma^*A}_{tot}$. 
The aim of this paper is to give a detailed derivation
of the formula for the total 
cross section.}

\maketitle

\section{Introduction}
\label{sec:1} 

The experimental observation  that the total $\gamma^*$-nucleus cross section 
at small Bj\/orken-$x$, $x_B$, is smaller than $A$ times the 
$\gamma^*$-nucleon cross section,
\beq
\sigma_{tot}^{\gamma^*A}<A\sigma_{tot}^{\gamma^*N},
\eeq 
is called shadowing.
Many theoretical efforts have been devoted to understand this phenomenon
quantitatively. A broad review of the experimental and theoretical situation can
be found in \cite{Arneodo}.
Depending on the reference frame, different physical pictures arise.
In the Breit frame, the nucleus appears contracted 
and parton fusion leads to a reduction of the parton density at
low Bj\/orken-$x$ \cite{kancheli}-\cite{q}. A very intuitive 
picture arises in the
rest frame of the nucleus, where shadowing may be 
understood qualitatively in the following
way: The virtual photon fluctuates into a hadronic state 
\cite{bauer}-\cite{kp} that
interacts with the nucleus at it's surface and  the nucleons
inside have less chances to interact with the photon. Thus, 
the total cross section
is smaller than expected naively.
 
When we want to describe shadowing in the rest frame of the nucleus, we have to
choose an appropriate basis in which the hadronic fluctuation is expanded. Since
the hadronic states must have the same quantum numbers as the photon, it is
reasonable to write the physical photon as a superposition of vector mesons.
This idea leads to the (generalized) vector meson dominance model (G)VMD. 
The fluctuation extends over a
distance called coherence length
\beq\label{coh}
L_c=\frac{2\nu}{Q^2+M^2_X},
\eeq
where $M_X$ is the invariant mass of the fluctuation, $\nu$ is the energy of
the photon and $Q^2$ it's virtuality. For small $x_B$, this length
can become much larger than the nuclear radius $R_A$. 
In this hadronic basis \cite{Gribov}, the shadowing correction is given by the
Karmanov-Kondratyuk-formula \cite{KKK}, i.~e.
\beqn\label{3K}
\frac{\sigma_{tot}^{\gamma^*A}}{A\sigma_{tot}^{\gamma^*N}} & = &
1-\frac{4\pi}{\sigma_{tot}^{\gamma^*N}}\left<T\right> \\\nonumber
& \times & \int d^2b\!
\int dM^2_X\,\frac{d^2\sigma\left(\gamma^*N\rightarrow XN\right)}
{dM_X^2\,dt}\Bigg|_{t=0}\!F_A^2\left(L_c,b\right), 
\eeqn
in the double scattering approximation.
For the case, $L_c\sim R_A$, the finite size of 
the nucleus has to be
taken into account. This is encoded in the nuclear form factor,
\beq
F_A^2\left(L_c,b\right)=\frac{1}{A\left<T\right>}\left|\,
\int_{-\infty}^{\infty}dz\, {n_A}\left(b,z\right)e^{iz/L_c}\right|^2,
\eeq
with
\beq
\left<T\right>=\frac{1}{A}\int d^2b\, T^2\left(b\right).
\eeq
Here, $T\left(b\right)=\inf
dz\,{n_A}\left(b,z\right)$ is the nuclear thickness, i.~e.~the integral of the
nuclear density over the direction of the incident photon
and $b$ is the impact parameter.
However, formula (\ref{3K}) 
takes only the double scattering term into account.
When we want to calculate corrections from higher order scattering terms, we are
faced with the problem,
that the vector mesons are not eigenstates of the interaction and
processes like $X N\rightarrow X^\prime N$, where the meson is scattered into
another state, are possible.

This problem may be solved by using the eigenstates of the interaction as
basis \cite{Zamo}. 
For very low Bjorken-$x$, $x_B<0.001$, 
such an eigenstate is a quark-antiquark 
pair with
fixed transverse separation $\mbox{\boldmath$\rho$}$. 
The separation is frozen during the propagation
through the nucleus, because of Lorentz time dilatation. 
At
higher values, $x_B\approx 0.01$ or $x_B\approx 0.1$, the finite size of the
nucleus will be important and we have to introduce the nuclear formfactor
$F_A\left(L_c,b\right)$.
This is a serious problem, because no systematic way is known to implement
$F_A\left(L_c,b\right)$ into higher order 
scattering terms than double scattering,
cmp.~eq.~(\ref{3K}). 
Even worse, we do not exactly know what $F_A$ is, because the coherence length,
eq.~(\ref{coh}), 
depends on the mass of the hadronic state $M_X$. This quantity is
well defined, when we use the hadronic basis, but for two quarks with fixed
transverse separation, no mass is defined.

A solution to these problems has been proposed 
in \cite{first} and numerical calculations
have shown the importance of multiple scattering, especially for heavy nuclei.
For lead it gives a correction about 50\%. However, the formula for
$\sigma_{tot}^{\gamma^*A}$ was given without derivation. Therefore, the aim of
this paper is to give a complete derivation and to point out the approximations.

Before we start with the derivation, we briefly sketch the idea of the approach.
When no higher Fock-states are taken into account, 
the total cross section $\sigma_{tot}^{\gamma^*A}$
is the cross section for production of a $q\bar q$-pair in the field of the
nucleus. This means, we calculate DIS from the elastic scattering of the 
$q\bar q$ component of the virtual photon off the nuclear target.
The total cross section for pair production on a single nucleon may be
written in the form \cite{NZ}
\beq
\sigma_{tot}^{\gamma^*N}=\int d\lambda\,\int d^2\rho\,
\left(\left|\bar\Phi_{T}\left(\varepsilon\mbox{\boldmath$\rho$}\right)\right|^2
+\left|\bar\Phi_{L}\left(\varepsilon\rho\right)\right|^2\right)
\sigma^N_{q\bar q}\left(\rho\right),
\eeq
with the total cross section $\sigma^N_{q\bar q}\left(\rho\right)$ 
for the scattering of the pair
off a nucleon. The probability for the virtual photon to fluctuate into a  
$q\bar q$-pair of transverse separation $\mbox{\boldmath$\rho$}$ is described 
by 
the transverse
and longitudinal light-cone
wavefunctions, summed over all flavors, colors and spin states
\beqn\label{LCWF}\nonumber
\left|\bar\Phi_{T}\left(\varepsilon\mbox{\boldmath$\rho$}\right)\right|^2&=&
\frac{6\alpha_{em}}{\left(2\pi\right)^2}\sum\limits_{f=1}^{N_f}Z_f^2
\left\{\left(1-2\lambda\left(1-\lambda\right)\right)\varepsilon^2
K_1\left(\varepsilon\rho\right)^2\right. \\ & & +\left.
m_f^2K_0\left(\varepsilon\rho\right)^2
\right\},\\
\left|\bar\Phi_{L}\left(\varepsilon\rho\right)\right|^2&=&
\frac{24\alpha_{em}}{\left(2\pi\right)^2}\sum\limits_{f=1}^{N_f}Z_f^2
Q^2\;\lambda^2\left(1-\lambda\right)^2K_0\left(\varepsilon\rho\right)^2.
\eeqn
Here, $Z_f$ is the flavor charge, $m_f$ the mass of a quark of flavor $f$,
$\alpha_{em}=1/137$ and $\varepsilon^2=\lambda\left(1-\lambda\right)Q^2+m_f^2$,
$\lambda$ is the light cone momentum fraction carried by the quark.
$K_0$ and $K_1$ are the MacDonald functions of zeroth and first order,
respectively.
We point out, that the pair is created electromagnetically
in a color singlett state, but interacts with
a nucleon via pomeron exchange.

For a nuclear target, the total cross section 
may be written in eikonal form \cite{NZ}
\beqn\label{eikonal}\nonumber
\sigma_{tot}^{\gamma^*A} & = & \int d\lambda\,\int d^2\rho\,
\left(\left|\bar\Phi_{T}\left(\varepsilon\mbox{\boldmath$\rho$}\right)\right|^2
+\left|\bar\Phi_{L}\left(\varepsilon\rho\right)\right|^2\right) \\
& \times &
2\int d^2b\,\left[1-\exp\left(-\frac{\sigma^N_{q\bar q}\left(\rho\right)}{2}
T\left(b\right)\right)\right],
\eeqn
if the transverse separation $\rho$ is frozen, i.~e.~at very small $x_B$.
In this approximation, one splits a fast oscillating phase factor from the 
wavefunction $\varphi(\vec r)=e^{ikz}\tilde\varphi(\vec r)$
and obtains for the slowly varying part $\tilde\varphi$
an equation of the form
\beq\label{SGL}
i\frac{\partial}{\partial z}\tilde\varphi=
\left(-\frac{\Delta_\perp}{const.}+V\right)\tilde\varphi.
\eeq 
In the target rest frame,
the interaction is given by a color-static potential $V$.
However, anticipating that all dependence of the interaction will be absorbed into the
dipole cross section $\sigma_{q\bar q}^N(\rho)$, we use
an abelian potential. Strictly speaking, this is justified only in the case of
an electron-positron pair propagating in a condensed medium, 
but our most important aim is to demonstrate
explicitely, how to treat fluctuations of the transverse size of the pair, for
values of $x_B$, where the transverse size is not yet frozen.
Since the influence of the potential will be expressed in terms of scattering
amplitudes and because
our final result interpolates between $A\sigma_{tot}^{\gamma^*N}$ and
eq.~(\ref{eikonal}), we assume, that our results hold also for the case
of a nonabelian potential and in the presence of inelastic processes. 

The Laplacian acts only on the transverse coordinates and is omitted in the
eikonal approximation.
Then, eq. (\ref{SGL}) is easily integrated.
The multiple scattering series is sum\-med like in Glauber theory \cite{Glauber}.
This is possible, because the typical distance between two nucleons inside a
nucleus is roughly $2$ fm, while the gluon correlation length is much smaller,
presumably $\sim 0.3$ fm. 
This approximation has been studied for the case of QCD by 
Mueller \cite{Al}
with the result that the scatterings off different
nucleons inside the nucleus are additive.
After averaging over the medium, one
obtains eq.~(\ref{eikonal}).

The idea is now to keep this Laplacian, because it
describes the transverse motion of the particles in the pair. 
Taking into account this motion, we can correctly describe the effective
mass of the fluctuation in a coordinate space representation.
The phase shift
function has then to be replaced by the Green function for eq.~(\ref{SGL}).
We write this Green function as a path integral and averaging over all
scattering centers yields an effective Green function $W$ with an absorbtive
optical potential 
$V_{opt}=-i\sigma^N_{q\bar q}\left(\rho\right)n_A\left(b,z\right)/2$.
All the details will be given in the next section.

To make the influence of multiple scattering as clear as possible, 
we represent
the total cross section in the form
\beq\label{form}
\sigma_{tot}^{\gamma^*A}=A\sigma_{tot}^{\gamma^*N}-\sigma_{tot}^{int},
\eeq
similar to eq.~(\ref{3K}),
where $\sigma_{tot}^{int}$ is an interference term that
accounts for 
multiple scattering. 
\begin{figure}[t] 
  \resizebox{0.48\textwidth}{!}{%
  \psfig{figure=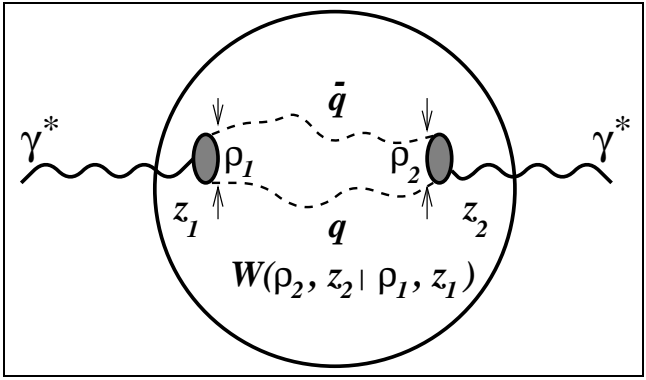} 
} \caption{A cartoon for the shadowing (negative)
term
in (\ref{form}). The Green function $W({\mbox{\boldmath$\rho$}}_2,
z_2|\mbox{\boldmath$\rho$}_1,z_1)$
results
from the summation over
different
paths of the $q\bar q$ pair propagation
through
the
nucleus.
} \label{event}
\end{figure}
This term, illustrated in fig. (\ref{event}), 
is due to destructive interference between pairs created at different 
coordinates $z$ and 
contains also the contribution of the
transverse motion of the pair to the coherence length.
It is 
described by the Green function  
$W$ and avoids the
appearance of the undefined quantity
$M_X$. The Green function also takes the finite size of 
the
nucleus into account.

\section{Derivation of the formula}
\label{sec:2} 

In order to derive explicit expressions for the terms in eq. (\ref{form}), we
start from the general expression for the cross section for the 
production of a
$q\bar q$-pair,
\beqn\label{ds}\nonumber
d\sigma^{\gamma^*A} & = & Z_f^2\alpha_{em}\,\left|M_{fi}\right|^2\,
\delta\left(\nu-p_{q}^0-p_{\bar{q}}^0\right)
\frac{d^3p_{q}\,d^3p_{\bar{q}}}{\left(2\pi\right)^4\nu\,2p_{q}^0\,
2p_{\bar{q}}^0}\\ \\
\label{ds2}
& = & Z_f^2\alpha_{em}\,\left|M_{fi}\right|^2\,
\frac{d^2p_{\perp,q}\,d^2p_{\perp,\bar{q}}\,d\lambda}{\left(2\pi\right)^4
4\nu^2\,
\lambda\left(1-\lambda\right)}
\eeqn
where $Z_f^2$ is the flavor charge, 
$\alpha_{em}=1/137$ and the matrix element is given by
\beq\label{matrix}
M_{fi}=\int d^3r\, \Psi_{q}^\dagger\left(\vec r\right)\,
\mbox{\boldmath${\alpha}$}\cdot\mbox{\boldmath$\epsilon$}
\;\Psi_{\bar{q}}\left(\vec r\right)
e^{ikz}.
\eeq
We choose the $z$-axis to lie in direction of propagation of the photon. The
photon's momentum is denoted by $k$ and it's energy by $\nu$, $p_q$ is
the momentum of the quark and $p_{\bar q}$ the momentum of the antiquark. 
In eq.~(\ref{ds2}) we have introduced the energy fraction $\lambda=p_q^0/\nu$.
In the
ultrarelativistic case we are considering, pair production takes place
predominantly in forward direction and therefore we distinguish between the
longitudinal direction ($z$-direction) and the transverse directions.
The longitudinal momenta are large compared to the flavor masses, $p_z\gg m_f$, 
and
to the perpendicular momenta, $\left|\vec p_{\perp}\right|^2\sim m_f^2$. In
our approximation, we keep terms of order $m_f/p_z$ only in exponentials
and neglect them otherwise. All higher
order terms are omitted.

Note, that the pair is created electromagnetically in a color singlet state,
therefore we have the factor $Z_f\alpha_{em}$ in eq.~(\ref{ds}), but 
we describe the
with the nucleons an abelian potential
$\phi\left(\vec r\right)$. 
The particles in the pair move in a potential 
$U\left(\vec r\right)$ that is a superposition of the
potentials of all nucleons,
\beq\label{super}
U\left(\vec r\right)=\sum\limits_{j=1}^A\phi\left(\vec r-\vec r_j\right).
\eeq
The vector $\vec r_j$ runs over all positions of the nucleons.

The 
wavefunction of the quark fullfills the Dirac equations and is an
eigenstate with positive energy, while the antiquark is represented by an
eigenstate with negative energy, 
\beqn
\left(p_{q}^0-U\left(\vec r\right)-m_f{\mbox{\boldmath$\beta$}}
+i\mbox{\boldmath$\alpha$}\cdot\mbox{\boldmath$\mbox{\boldmath$\nabla$}$}
\right)\Psi_{q}\left(\vec r\right) & = & 0,\\
\left(-p_{\bar{q}}^0-U\left(\vec r\right)-m_f{\mbox{\boldmath$\beta$}}
+i\mbox{\boldmath$\alpha$}\cdot\mbox{\boldmath$\mbox{\boldmath$\nabla$}$}\right)
\Psi_{\bar{q}}\left(\vec r\right) & = & 0.
\eeqn
No interaction between the quark and the antiquark is taken into account and
therefore, the two equations decouple.
The wavefunction of the quark $\Psi_{q}\left(\vec r\right)$ 
contains an outgoing plane wave and an
outgoing spherical wave in it's asymptotic
form, while the wavefunction of the antiquark contains an incoming  
spherical wave and an incoming plane wave. 
We transform these equations into second order
equations by applying the operator 
$\left(p_{q}^0-U\left(\vec r\right)\right.$ $\left.+
m_f{\mbox{\boldmath$\beta$}}
-i\mbox{\boldmath$\alpha$}\cdot\mbox{\boldmath$\mbox{\boldmath$\nabla$}$}\right)$ 
on the first 
and the
cor\-re\-spond\-ing operator on the second equation, as described in \cite{LL}.
 When we omit the term quadratic
in the potential, we obtain
\beqn
\left(\Delta+\left|\vec p_{q}\right|^2\!-2p_{q}^0U\!\!\left(\vec r\right)
+i\mbox{\boldmath$\alpha$}\cdot\left(\mbox{\boldmath$\mbox{\boldmath$\nabla$}$} 
U\!\!\left(\vec r\right)\right)
\right)\Psi_{q}\left(\vec r\right) & = & 0, \\
\left(\Delta+\left|\vec p_{\bar{q}}\right|^2\!+
2p_{\bar{q}}^0U\!\!\left(\vec r\right)
+i\mbox{\boldmath$\alpha$}\cdot\left(\mbox{\boldmath$\mbox{\boldmath$\nabla$}$} 
U\!\!\left(\vec r\right)\right)
\right)\Psi_{\bar{q}}\left(\vec r\right) & = & 0. 
\eeqn
The solutions may approximately 
be written as Furry-Sommerfeld-Maue 
\cite{Furry,SM}
type
wavefunctions,
\beqn\label{sm1}
\Psi_{q}\left(\vec r\right) & = &
e^{i\vec p_{q}\cdot\vec r}
\left(1-\frac{i\mbox{\boldmath$\alpha$}}{2p_{q}^0}\cdot\mbox{\boldmath$\mbox{\boldmath$\nabla$}$}
\right)\,
F_{q}\left(\vec r\right)\,
u\left(p_{q},\lambda_{q}\right), \\
\label{sm2}
\Psi_{\bar q}\left(\vec r\right) & = &
e^{-i\vec p_{q}\cdot\vec r}
\left(1+\frac{i\mbox{\boldmath$\alpha$}}{2p_{\bar q}^0}\cdot
\mbox{\boldmath$\mbox{\boldmath$\nabla$}$}
\right)\,
F_{\bar q}\left(\vec r\right)\,
v\left(p_{\bar q},\lambda_{\bar q}\right).
\eeqn
Here, $u\left(p_{q},\lambda_{q}\right)$ is the free spinor with positive energy
and pol\-ari\-zation $\lambda_q$. It satisfies
$\left(p_q\!\!\!\!\!/-m_f\right)u\left(p_{q},\lambda_{q}\right)=0$ and similarly
$\left(p_{\bar q}\!\!\!\!\!/+m_f\right)v\left(p_{q},\lambda_{q}\right)=0$. 
In Dirac representation they read
\beqn\label{Dirac1}
u\left(p_{q},\lambda_{q}\right) & = &
\sqrt{p^0_q+m_f}\left(
\begin{array}{c}
\displaystyle{\chi_q } \\
\displaystyle{\frac{\mbox{\boldmath$\sigma$}\cdot\vec p_q}{p^0_q+m_f}\;\chi_q} 
\end{array}\right),
\\
\label{Dirac2}v\left(p_{q},\lambda_{q}\right) & = &
\sqrt{p^0_{\bar q}+m_f}\left(
\begin{array}{c}
\displaystyle{\frac{\mbox{\boldmath$\sigma$}\cdot\vec p_{\bar{q}}}
{p^0_{\bar q}+m_f}\;
\chi_{\bar q}} \\ 
\displaystyle{\chi_{\bar q} }
\end{array}\right).
\eeqn
The three Pauli spin matrices are denoted by $\mbox{\boldmath$\sigma$}$ and the Pauli spin
state referred to the rest frame of the particle is $\chi_q$, or
$\chi_{\bar q}$ respectively. This means explicitely 
$\vec s\cdot\mbox{\boldmath$\sigma$}\,\chi_q=\lambda_{q}\chi_q$ and 
$\vec s\cdot\mbox{\boldmath$\sigma$}\,\chi_{\bar q}=-\lambda_{\bar q}\chi_{\bar q}$ 
with a spin vector $\vec
s$ normalized to unity and $\lambda_q,\lambda_{\bar q}=\pm 1$.
The
functions $F_{q}$ and $F_{\bar q}$ have no spinor structure any more.
They contain all dependence of the potential and have to be calculated for a
given $U\left(\vec r\right)$.
They fullfill the equations \cite{LL}
\beqn
\label{erste}
\left(\Delta
+2i\vec p_{q}\cdot\mbox{\boldmath$\nabla$}
-2p_{q}^0U\left(\vec r\right)
\right)F_{q}\left(\vec r\right)
& = & 0,\\
\label{zweite}\left(\Delta
-2i\vec p_{\bar q}\cdot\mbox{\boldmath$\nabla$}
+2p_{\bar q}^0U\left(\vec r\right)
\right)F_{\bar q}\left(\vec r\right) & = & 0,
\eeqn
with boundary conditions $F\rightarrow 1$ for the quark and for the antiquark as
$z \rightarrow \infty$. Note, that it is essential to take the correction
proportional to $\mbox{\boldmath$\alpha$}$ in eq.~(\ref{sm1}) and (\ref{sm2}) into account,
although these terms seem to be suppressed by a factor of $1/p^0$. It turns out,
that when we calculate the matrix element of the current operator,
$\left(\mbox{\boldmath$\alpha$}\cdot\mbox{\boldmath$\epsilon$}\,\right)_{fi}$, between free spinors,
 the large part cancels and we
are left with a contribution of the same order as produced by the correction
term. Thus, the terms proportional to $\mbox{\boldmath$\alpha$}$ may not be neglected in the
matrix element, although they give small corrections to the wave functions.

In order to remove the dependence on the transverse momenta from the phase
factors in eq.~(\ref{sm1}) and (\ref{sm2}), we rewrite
the solutions in the  form
\beqn\label{ansatz}\nonumber
\Psi_{q}\left(\vec r\right) & = &
e^{i \left|\vec p_q\,\right|z}
\left(1-\frac{i\vec{\mbox{\boldmath$\alpha$}}}{2p_{q}^0}\cdot\mbox{\boldmath$\nabla$}
-\frac{\vec{\mbox{\boldmath$\alpha$}}_\perp}{2p_{q}^0}\cdot\vec p_{\perp,q}\right.\\ & &
+\left.\frac{{\mbox{\boldmath$\alpha$}}_z}{2p_{q}^0}
\left(\left|\vec p_{q}\,\right|-p_{z,q}\right)
\right)\,
\psi_{q}\left(\vec r\right)\,
u\left(p_{q},\lambda_{q}\right), \\
\label{ansatz2}\nonumber
\Psi_{\bar q}\left(\vec r\right) & = &
e^{-i \left|\vec p_{\bar q}\,\right|z}
\left(1+\frac{i\vec{\mbox{\boldmath$\alpha$}}}{2p_{\bar q}^0}\cdot\mbox{\boldmath$\nabla$}
-\frac{\vec{\mbox{\boldmath$\alpha$}}_\perp}{2p_{\bar{q}}^0}
\cdot\vec p_{\perp,\bar{q}}\right.\\ & &
\left.+\frac{{\mbox{\boldmath$\alpha$}}_z}{2p_{q}^0}
\left(\left|\vec p_{\bar q}\,\right|-p_{z,\bar q}\right)
\right)\,
\psi_{\bar q}\left(\vec r\right)\,
v\left(p_{\bar q},\lambda_{\bar q}\right),
\eeqn
with
\beqn\label{psidef1}
\psi_{q}\left(\vec r\right) & = &
e^{i\vec p_{\perp,q}\cdot\vec r_\perp}
e^{-i\left(\left|\vec p_{ q}\,\right|-p_{z,q}\right)z}
F_{q}\left(\vec r\right), \\
\label{psidef2}\psi_{\bar q}\left(\vec r\right) & = &
e^{-i\vec p_{\perp,\bar{q}}\cdot\vec r_\perp}
e^{i\left(\left|\vec p_{\bar q}\,\right|-p_{z,\bar q}\right)z}
F_{\bar q}\left(\vec r\right),
\eeqn
and $
\left|\vec p\,\right|=\sqrt{p_0^2-m_f^2}$
for the quark and the antiquark respectively. In 
the following, we neglect the terms containing 
${\mbox{\boldmath$\alpha$}}_z$ in
eq.~(\ref{ansatz}) 
and (\ref{ansatz2}), because they are of order $O\left(1/p^0_qp^0_{\bar
q}\right)$. The functions $\psi_{q}\left(\vec r\right)$ and 
$\psi_{\bar q}\left(\vec r\right)$ will play the role of effective wave
functions for the quarks.

The phase factors 
combine in the matrix element (\ref{matrix}) 
to the minimal longitudinal momentum transfer, 
\beq
q_L^{min}=
k-\left|\vec p_{\bar q}\,\right|-\left|\vec p_{\bar q}\,\right|
\approx\frac{Q^2}{2\nu}
+\frac{m_f^2}{2p_{0,q}}
+\frac{m_f^2}{2p_{0,\bar{q}}},
\eeq
and we obtain
\beqn\nonumber
\lefteqn{M_{fi} \; =\;\int d^3 r e^{iq_L^{min}z} }\\
\nonumber &\times& 
\left[u^\dagger\left(p_{q},\lambda_{q}\right)
\left(\,\mbox{\boldmath$\alpha$}\cdot\mbox{\boldmath$\epsilon$}\;
+\mbox{\boldmath$\alpha$}\cdot
\frac{i\mbox{\boldmath$\nabla$}\left(\vec r_{q}\right)-
\vec{p}_{\perp,q}}{2p_q^0}\,
\mbox{\boldmath$\alpha$}\cdot\mbox{\boldmath$\epsilon$}\;\right.\right. \\ \nonumber & &
\left.\left.+\mbox{\boldmath$\alpha$}\cdot\mbox{\boldmath$\epsilon$}\,\vec{\bf \alpha}\cdot
\frac{i\mbox{\boldmath$\nabla$}\left(\vec r_{\bar{q}}\right)
-\vec{p}_{\perp,\bar{q}}}{2p_{\bar{q}}^0}
\right)
v\left(p_{\bar{q}},\lambda_{\bar{q}}\right)\right] \\
& \times &
\psi_{q}^*\left(\vec r_{{q}}\right)
\psi_{\bar{q}}\left(\vec r_{\bar{q}}\right)
\Big|_{\vec r=\vec r_{{q}}=\vec r_{\bar{q}}}.
\eeqn
Here, the coherence length, $L_c^{max}=1/q_L^{min}$, comes into the game as an
oscillating phase factor. However, $L_c^{max}$ does not depend on the
transverse momenta. Their influence on the cross section is encoded in the
rest of the wave functions, eq.~(\ref{ansatz}) and (\ref{ansatz2}).
The operator $\mbox{\boldmath$\nabla$}\left(\vec r_{q}\right)$ acts only on the variable
$\vec r_{q}$ and the operator 
$\mbox{\boldmath$\nabla$}\left(\vec r_{\bar{q}}\right)$ only on
$\vec r_{\bar q}$.
After the derivatives have been performed, the whole integrand 
has to be evaluated at 
$\vec r=\vec r_{{q}}=\vec r_{\bar{q}}$.

With the representation (\ref{Dirac1}) , (\ref{Dirac2})  
we obtain after some algebra
within the demanded accuracy
\beqn\nonumber\label{me}
\lefteqn{M_{fi} \; =\;\inf dz e^{iq_L^{min}z} \int d^2 r_\perp
\frac{1}{\sqrt{\lambda\left(1-\lambda\right)}}}
\\
\nonumber &\times& \chi_q^\dagger
\displaystyle{\Bigg\{ }
m_f\mbox{\boldmath$\sigma$}\cdot\mbox{\boldmath$\epsilon$}_T \\ & & \nonumber
+i\left(1-\lambda\right)\mbox{\boldmath$\sigma$}\cdot\vec e_z
\mbox{\boldmath$\epsilon$}_T\cdot\mbox{\boldmath$\nabla$}
\left(\vec r_{\perp,\bar{q}}\right)
+i\lambda\mbox{\boldmath$\sigma$}\cdot\vec e_z
\mbox{\boldmath$\epsilon$}_T\cdot\mbox{\boldmath$\nabla$}
\left(\vec r_{\perp,{q}}\right)\\[1.5ex]
\nonumber & & 
+\left(1-\lambda\right)\left(\vec e_z\times\mbox{\boldmath$\epsilon$}_T\right)
\cdot\mbox{\boldmath$\nabla$}\left(\vec r_{\perp,{\bar q}}\right)
-\lambda\left(\vec e_z\times\mbox{\boldmath$\epsilon$}_T\right)
\cdot\mbox{\boldmath$\nabla$}\left(\vec r_{\perp,{q}}\right)\\
\nonumber & & 
+\;2\,Q\,\lambda\left(1-\lambda\right)
\Bigg\}\chi_{\bar q}
 \\
& \times &
\psi_{q}^*\left(\vec r_{\perp,{q}},z\right)
\psi_{\bar{q}}\left(\vec r_{\perp,\bar{q}},z\right)
\Big|_{\vec r_{\perp,{q}}=\vec r_{\perp,\bar{q}}}.
\eeqn
Some details of the calculation can be found in \cite{OM}. The unit vector in
$z$-direction is denoted by $\vec e_z$.
The polarisation vector $\mbox{\boldmath$\epsilon$}_T$ corresponds to transverse
states of the $\gamma^*$, while the last term in the curly brackets
is due to longitudinal polarisation.
There are two
remarkable aspect concerning this last equation. First, it does not contain any
derivative with respect to $z$ any more, because of the transverse nature of the
polarization vector $\mbox{\boldmath$\epsilon$}_T$ and of $\vec e_z\times\mbox{\boldmath$\epsilon$}_T$. 
Second, all dependence on the transverse momenta in the spinor part 
has cancelled.

From the eq.~(\ref{erste}) and (\ref{zweite}) and from the definition of $\psi$,
eq.~(\ref{psidef1}) and (\ref{psidef2}), we can obtain an equation for $\psi$.
Assuming that these functions are only slowly 
varying with $z$, 
we omit the longitudinal part of the Laplacian and obtain
\beqn\label{Schroe1}
i\frac{\partial}{\partial z}
\psi_{q}\left(\vec r_\perp,z\right)
 & = & 
\left(-\frac{\Delta_\perp}{2p_{{q}}^0}
+U\left(\vec r_\perp,z\right)\right)
\psi_{q}\left(\vec r_\perp,z\right), \\
\label{Schroe2}
i\frac{\partial}{\partial z}
\psi_{\bar{q}}\left(\vec r_\perp,z\right)
& = & 
\left(\frac{\Delta_\perp}{2p_{\bar{q}}^0}
+U\left(\vec r_\perp,z\right)\right)
\psi_{\bar{q}}\left(\vec r_\perp,z\right).
\eeqn
Our ansatz yields two dimensional Schr\"odinger equations, where the
$z$-coordinate plays the role of time and the mass is given by the
energy. 
The Laplacian $\Delta_\perp$ acts on the transverse
coordinates only. The functions $\psi_{q}\left(\vec r_\perp,z\right)$ 
and $\psi_{\bar q}\left(\vec r_\perp,z\right)$ 
become two dimensional plane waves for $z\rightarrow \infty$, up to a phase
factor that cancels in the square of the matrix element.
It should be mentioned that the kinetic energy for the
antiquark has a negative sign.
This results from the fact that solutions of the Dirac equation with negative
energy propagate backwards in time.
The Laplacian in (\ref{Schroe1}) and (\ref{Schroe2}), 
account for the transverse motion of the pair in which we are
especially interested.
The functions $\psi_{\bar q}\left(\vec r_\perp,z\right)$ and
$\psi_{q}\left(\vec r_\perp,z\right)$ in the matrix element (\ref{me})
may now be expressed in terms of the Green functions for 
(\ref{Schroe1}) and (\ref{Schroe2}) and its asymptotic behaviour,
\beqn\nonumber
\psi_{q}\left(\vec r_{\perp,2},z_2\right) & = &
\int d^2 r_{\perp,1}\,G_{q}\left(\vec r_{\perp,2},z_2\,
|\,\vec r_{\perp,1},z_\infty\right)\\ & &
\times\;e^{i\vec p_{\perp,q}\cdot\vec r_{\perp,1}}
e^{-i\left(\left|\vec p_{ q}\,\right|-p_{z,q}\right)z_\infty},\\
\nonumber\psi_{\bar q}\left(\vec r_{\perp,2},z_2\right) & = &
\int d^2 r_{\perp,1}\,G_{\bar q}\left(\vec r_{\perp,2},z_2\,
|\,\vec r_{\perp,1},z_\infty\right)\\ & &
\times\;e^{-i\vec p_{\perp,\bar q}\cdot\vec r_{\perp,1}}
e^{i\left(\left|\vec p_{\bar q}\,\right|-p_{z,\bar q}\right)z_\infty}.
\eeqn

Now we use the
expression for $d\sigma^{\gamma^*A}$, eq.~(\ref{ds}), 
and with the matrix element (\ref{me})
and the last two relations, we obtain
\beqn
\nonumber d\sigma^{\gamma^*A} & = & Z_f^2\alpha_{em}
\;\inf dz \inf dz^\prime e^{iq_L^{min}\left(z-z^\prime\right)} \\ & &
\nonumber \int d^2 r_\perp\,d^2\tau_{q}\,d^2\tau_{\bar{q}}\,
\int d^2 r_\perp^{\,\prime}\,d^2\tau_{q}^\prime\,d^2\tau_{\bar{q}}^\prime\, \\
\nonumber & &\;\;\;\times\;\;
{\cal O}\left(\vec r_{\perp,{q}},\vec r_{\perp,\bar{q}}\right)
G_{q}^*\left(\vec r_{\perp,{q}},z\,|\,\mbox{\boldmath$\tau$}_{q},z_{\infty}\right)\,\\
\nonumber & & \qquad\qquad
G_{\bar{q}}\left(\vec r_{\perp,\bar{q}},z\,
|\,\mbox{\boldmath$\tau$}_{\bar{q}},z_{\infty}\right)
\Big|_{\vec r_{\perp}=\vec r_{\perp,{q}}=\vec r_{\perp,\bar{q}}}
\\
\nonumber & &\;\;\;\times\;\;
{\cal O}^*\left(\vec r^{\,\prime}_{\perp,{q}}
,\vec r^{\,\prime}_{\perp,\bar{q}}\right)
G_{q}\left(\vec r_{\perp,{q}}^{\,\prime},z^{\,\prime}\,
|\,\mbox{\boldmath$\tau$}_{q}^{\,\prime},z_{\infty}\right)\,\\
\nonumber & & \qquad\qquad
G_{\bar{q}}^*\left(\vec r_{\perp,\bar{q}}^{\,\prime},z^{\,\prime}\,
|\,\mbox{\boldmath$\tau$}_{\bar{q}}^{\,\prime},z_{\infty}\right)
\Big|_{\vec r_{\perp}^{\,\prime}=\vec r_{\perp,{q}}^{\,\prime}
=\vec r^{\,\prime}_{\perp,\bar{q}}}
\\
\nonumber & &\;\;\;\times\;\;
e^{i\left(\mbox{\boldmath$\tau$}_{q}^{\,\prime}-\mbox{\boldmath$\tau$}_{q}\right)\cdot\vec p_{\perp,q}}
\,e^{i\left(\mbox{\boldmath$\tau$}_{\bar{q}}^{\,\prime}-\mbox{\boldmath$\tau$}_{\bar{q}}\right)
\cdot\vec p_{\perp,\bar{q}}}\\
& &\;\;\;\times\;\;
\frac{d^2p_{\perp,q}\,d^2p_{\perp,\bar{q}}\,d\lambda}
{\left(2\pi\right)^4\,\lambda^2\left(1-\lambda\right)^2 4\nu^2}.
\eeqn
For convenience we have introduced the operator
\beqn\nonumber
\lefteqn{{\cal O}\left(\vec r_{\perp,{q}}
,\vec r_{\perp,\bar{q}}\right)}\\ & = &\nonumber
\chi_q^\dagger
\displaystyle{\Bigg\{ }
m_f\mbox{\boldmath$\sigma$}\cdot\mbox{\boldmath$\epsilon$}_T\\ & + &\nonumber
i\left(1-\lambda\right)\mbox{\boldmath$\sigma$}\cdot\vec e_z
\mbox{\boldmath$\epsilon$}_T\cdot\mbox{\boldmath$\nabla$}
\left(\vec r_{\perp,\bar{q}}\right)
+i\lambda\mbox{\boldmath$\sigma$}\cdot\vec e_z
\mbox{\boldmath$\epsilon$}_T\cdot\mbox{\boldmath$\nabla$}
\left(\vec r_{\perp,{q}}\right)\\[1.5ex]
 & + & \nonumber
\left(1-\lambda\right)\left(\vec e_z\times\mbox{\boldmath$\epsilon$}_T\right)
\cdot\mbox{\boldmath$\nabla$}\left(\vec r_{\perp,{\bar q}}\right)
-\lambda\left(\vec e_z\times\mbox{\boldmath$\epsilon$}_T\right)
\cdot\mbox{\boldmath$\nabla$}\left(\vec r_{\perp,{q}}\right)\\
 & + & 
2\,Q\,\lambda\left(1-\lambda\right)
\Bigg\}\chi_{\bar q}.
\eeqn
In order to obtain the total cross section, we integrate over 
$\vec p_{\perp,q}$ and
$\vec p_{\perp,\bar{q}}$. The exponential factors give a
$\delta$-function that enables us to perform the integrations over all the
\boldmath$\tau$\unboldmath s. 
Note, that ${\cal O}$ does not depend on \boldmath$\tau$\unboldmath. 
We get
\beqn
\nonumber \sigma_{tot}^{\gamma^*A} & = & Z_f^2\,
\frac{\alpha_{em}}{\nu^2}\,2\Re\int\limits_{0}^1 
\frac{d\lambda}{\lambda^2\left(1-\lambda\right)^2}\\ 
\nonumber &\times &
\;\int\limits_{-\infty}^{\infty} dz \int\limits_{z}^{\infty} dz^\prime 
e^{iq_L^{min}\left(z-z^\prime\right)}
\int d^2 r_\perp\,\int d^2 r_\perp^{\,\prime} 
 \\ 
\nonumber &\times &
{\cal O}\left(\vec r_{\perp,{q}},\vec r_{\perp,\bar{q}}\right)
{\cal O}^*\left(\vec r^{\,\prime}_{\perp,{q}}
,\vec r^{\,\prime}_{\perp,\bar{q}}\right)
\\[1.2ex]
\label{stot} &\times &
G_{q}\left(\vec r_{\perp,q}^{\,\prime},z^{\,\prime}\,
|\vec r_{\perp,q},z\right)\,\\ 
&\times &\nonumber 
G_{\bar{q}}^*\left(\vec r_{\perp,\bar q}^{\,\prime},z^{\,\prime}\,
|\vec r_{\perp,\bar q},z\right)\!\!
\Big|_{\vec r_{\perp}=\vec r_{\perp,{q}}=\vec r_{\perp,\bar{q}}\,;\,
\vec r_{\perp}^{\,\prime}=\vec r_{\perp,{q}}^{\,\prime}
=\vec r^{\,\prime}_{\perp,\bar{q}}}.
\eeqn
Instead of four propagators we are left with only two, because we have used the
convolution relation
\beqn\nonumber
\lefteqn{G\left(\vec r_{\perp,2},z_2\,|\,\vec r_{\perp,1},z_1\right)}\\
& =&
\int d^2 r_\perp\;
G\left(\vec r_{\perp,2},z_2\,|\,\vec r_\perp,z\right)
G\left(\vec r_\perp,z\,|\,\vec r_{\perp,1},z_1\right).
\eeqn

In order to derive an expression that is convenient for numerical calculations,
we make use of the path-integral 
 representation of the propagators in eq.~(\ref{stot}). They read 
\beqn\label{path} \nonumber
\lefteqn{G\left(\vec r_{\perp}^{\,\prime},z^{\,\prime}\,
|\vec r_{\perp},z\right)}\\
& = & \!\!\int{\cal D}\mbox{\boldmath$\tau$}\,\exp\left\{
i\int\limits^{z^{\,\prime}}_{z}d\xi\left(\pm\frac{p^0}{2}
\dot{\mbox{\boldmath$\tau$}}^{\,2}-
U\left(\mbox{\boldmath$\tau$},\xi\right)\right)\right\}\\
 \nonumber\label{path2}& 
 \approx &\!\! \int{\cal D}\mbox{\boldmath$\tau$}\,\exp\left\{
i\!\int\limits^{z^{\,\prime}}_{z}\pm \frac{p^0}{2}
\dot{\mbox{\boldmath$\tau$}}^{\,2}d\xi\right.
\\ \nonumber &&\left.
-i
\sum\limits_{j=1}^A\mbox{X}\left(\mbox{\boldmath$\tau$}\left(z_j\right)-
\vec r_{j,\perp}\right)
\vartheta\left(z^{\,\prime}-z_j\right)\vartheta\left(z_j-z\right)
\right\},\nonumber \\
\eeqn
where the upper sign corresponds to the quark and the lower to the antiquark.
In this expression, $\mbox{\boldmath$\tau$}$ is a function of $\xi$. 
The derivative with
respect to $\xi$ is denoted by $\dot{\mbox{\boldmath$\tau$}}$.
Obviously, the condition
\beq
G\left(\vec r_{\perp}^{\,\prime},z^{\,\prime}\,
|\vec r_{\perp},z\right)\Big|_{z^{\,\prime}=z}
=\delta^{(2)}\left(\vec r_{\perp}^{\,\prime}-\vec r_{\perp}\right)
\eeq
has to be fullfilled and further it must obtain
$\mbox{\boldmath$\tau$}\left(z\right)=\vec r_{\perp}$ and
$\mbox{\boldmath$\tau$}\left(z^{\,\prime}\right)=\vec r_{\perp}^{\,\prime}$.
In eq.~(\ref{path2}), we have introduced the phase shift function 
$\mbox{X}\left(\mbox{\boldmath$\tau$}-\vec r_{j,\perp}\right)$. 
The step function 
$\vartheta\left(z\right)$ is $0$ for $z<0$ and $1$ for $z>0$.
As
mentioned before, $U$ is the superposition of all potentials 
of the nucleons, see eq.~(\ref{super}). Let
$\phi$ be the potential of a single nucleon in the target. 
The position of the nucleon with number $j$ is denoted by the transverse vector
$\vec r_{j,\perp}$ and the longitudinal coordinate $z_j$.
If the range
of interaction is much smaller than the distance 
$z^{\,\prime}-{z}$, the potential is practically zero outside the domain of
integration over $\xi$ and the phase shift function 
is  given by
\beq
\mbox{X}\left(\vec r_\perp-\vec r_{j,\perp}\right)
=\inf
d\xi
\,\phi\left(\vec r_\perp-\vec r_{j,\perp},\xi-z_j\right).
\eeq
We have replaced $\mbox{\boldmath$\tau$}\left(\xi\right)$ by 
$\mbox{\boldmath$\tau$}\left(z_j\right)$. This means,
 we use only an average
value of the transverse coordinate for calculating the phase shift for
scattering off a single nucleon.

The two path integrals sum over all possible trajectories of the two particles.
In order to calculate the cross section for pair production in the nuclear
medium, we have to average over all nucleons.
We obtain with the path-integral (\ref{path2})
\beqn
\nonumber
\lefteqn{
\biggl<G_{q}\left(\vec r_{\perp,q}^{\,\prime},z^{\,\prime}\,
|\vec r_{\perp,q},z\right)\,
G_{\bar{q}}^*\left(\vec r_{\perp,\bar q}^{\,\prime},z^{\,\prime}\,
|\vec r_{\perp,\bar q},z\right)\biggr>}
\\
& = &\nonumber\left< \int{\cal D}\mbox{\boldmath$\tau$}_q\,\int{\cal D}
\mbox{\boldmath$\tau$}_{\bar q}\,
\exp\Biggl\{
i\int\limits^{z^{\,\prime}}_{z}\left(
\frac{p_{q}^0}{2}\dot{\mbox{\boldmath$\tau$}}_q^{\,2}
+\frac{p^0_{\bar{q}}}{2}\dot{\mbox{\boldmath$\tau$}}_{\bar q}^{\, 2}\right)d\xi
\Biggr.\right.\\
& & \nonumber
+i
\sum\limits_{j=1}^A
\vartheta\left(z^{\,\prime}-z_j\right)\vartheta\left(z_j-z\right)\\
& & \left. \Biggl. \Bigl(
\mbox{X}\left(\mbox{\boldmath$\tau$}_{\bar q}\left(z_j\right)
-\vec r_{j,\perp}\right)
-\mbox{X}\left(\mbox{\boldmath$\tau$}_{q}\left(z_j\right)
-\vec r_{j,\perp}\right)
\Bigr)\Biggr\}\right>,
\eeqn
with boundary conditions 
$\mbox{\boldmath$\tau$}_q\left(z\right)=\vec r_{\perp,q}$,
$\mbox{\boldmath$\tau$}_q\left(z^{\,\prime}\right)=\vec r_{\perp,q}^{\,\prime}$,
$\mbox{\boldmath$\tau$}_{\bar q}\left(z\right)=\vec r_{\perp, \bar q}$ and
$\mbox{\boldmath$\tau$}_{\bar q}\left(z^{\,\prime}\right)
=\vec r_{\perp,\bar q}^{\,\prime}$.
The averaging procedure is similar to the one described in \cite{Glauber}.
We neglect all correlations between the nucleons and introduce the 
average
nuclear density ${n_A}$, which is normalized to $A$,
 eq. (\ref{density}). Then, the whole expression may be
written as an exponential, if $A$ is large enough,
\beqn
\nonumber\lefteqn{
\left<\exp\left(i
\sum\limits_{j=1}^A
\vartheta\left(z^{\,\prime}-z_j\right)\vartheta\left(z_j-z\right)
\right.\right.}
\\\nonumber&&
\Biggl.\Biggl.\Bigl(
\mbox{X}\left(\mbox{\boldmath$\tau$}_{\bar q}\left(z_j\right)-
\vec r_{j,\perp}\right)
-\mbox{X}\left(\mbox{\boldmath$\tau$}_{q}\left(z_j\right)-
\vec r_{j,\perp}\right)
\Bigr)
\Biggr)\Biggr>\\\nonumber
\label{density}& = & \Biggl\{1-\frac{1}{A}\int d^2s
\int\limits^{z^{\,\prime}}_{z}d\xi^{\,\prime}
{n_A}\left(\vec s,\xi^{\,\prime}\right)\Biggl(1\Biggr.\Biggr.\\
&&\left.\left.
-\exp\Biggl(i
\Bigl(
\,\mbox{X}\left(\mbox{\boldmath$\tau$}_{\bar q}\left(\xi^{\,\prime}\right)-\vec s\right)
-\,\mbox{X}\left(\mbox{\boldmath$\tau$}_q\left(\xi^{\,\prime}\right)-\vec s\right)
\Bigr)\Biggr)\right)
\right\}^A
\\
\label{app}
& \approx &
\exp\Biggl\{-\int\limits^{z^{\,\prime}}_{z} d\xi^{\,\prime}
{n_A}\left(b,\xi^{\,\prime}\right)\int d^2s\Biggr.\\
\nonumber&\times &\left.
\left(1-
\exp\Biggl(i
\Bigl(
\,\mbox{X}\left(\mbox{\boldmath$\tau$}_{\bar q}\left(\xi^{\,\prime}\right)
-\vec s\right)
-\,\mbox{X}\left(\mbox{\boldmath$\tau$}_q\left(\xi^{\,\prime}\right)
-\vec s\right)\Bigr)\Biggr)
\right)\right\}.
\eeqn
When 
${n_A}$ is varying
fairly smoothly inside the nucleus, 
we can replace its dependence of $\vec s$, the transverse distance between the
pair and the scattering nucleon, by the impact parameter $\vec b$,
eq.~(\ref{app}).
Since we have a short ranged interaction, it is reasonable to choose as impact
parameter $\vec b=\left(\vec r_\perp+\vec r_\perp^{\,\prime}\right)/2$.
This way we find the forward scattering amplitude 
for a dipole scattering off a single
nucleon, see eq.~(\ref{app}). 
The corresponding cross section,
\beq\label{ff}
\sigma_{q\bar{q}}^N
\left(\rho\right)
=2\Re \int d^2s
\left(1-\exp\Biggl(i
\Bigl(
\mbox{X}\left(\mbox{\boldmath$\rho$}-\vec s\right)
-\mbox{X}\left(\vec s\right)\Bigr)\Biggr)\right),
\eeq
 appears as imaginary
potential in the propagator.
Omitting the realpart of the  ampli\-tude, 
which is known to be small,
we finally arrive at
\beqn
\nonumber
\lefteqn{
\biggl<G_{q}\left(\vec r_{\perp,q}^{\,\prime},z^{\,\prime}\,
|\vec r_{\perp,q},z\right)\,
G_{\bar{q}}^*\left(\vec r_{\perp,\bar q}^{\,\prime},z^{\,\prime}\,
|\vec r_{\perp,\bar q},z\right)\biggr>}
\\
& = &
\int{\cal D}\mbox{\boldmath$\tau$}_q\,
\int{\cal D}\mbox{\boldmath$\tau$}_{\bar q}\,
\exp\Biggl\{
i\int\limits^{z^{\,\prime}}_{z}d\xi\Biggl(
\frac{p_{q}^0}{2}\dot{\mbox{\boldmath$\tau$}}_q^{\,2}
+\frac{p^0_{\bar{q}}}{2}\dot{\mbox{\boldmath$\tau$}}_{\bar q}^{\, 2}
\Biggr.\Biggr.
\\&&\nonumber
\qquad\qquad\qquad\qquad
+\Biggl.\left.
\frac{i}{2}{n_A}\left(b,\xi\right)\sigma_{q\bar{q}}^N
\left(\left|\mbox{\boldmath$\tau$}_q
-\mbox{\boldmath$\tau$}_{\bar q}\right|\right)
\right)\Biggr\}.
\eeqn
Note, that all dependence on the potential of the nucleons,
$\phi\left(\vec r\right)$,
has been absorbed into $\sigma_{q\bar{q}}^N$. 
Although eq.~(\ref{ff}) looks different for a nonabelian potential,
we belief, that 
the formulae presented in this paper still hold, if
$\sigma_{q\bar{q}}^N$ is taken from experiment or calculated
in perturbative QCD.

It is
convenient to introduce center of mass coordinates, 
$\mbox{\boldmath$\tau$}_{rel}=\mbox{\boldmath$\tau$}_{\bar q}
-\mbox{\boldmath$\tau$}_q$ and
$\mbox{\boldmath$\tau$}_{cm}=\left(1-\lambda\right)
\mbox{\boldmath$\tau$}_{\bar q}+\lambda\mbox{\boldmath$\tau$}_q$, 
and to
express the sum of the two kinetic energies
as the sum of the kinetic energy of the relative motion
and the center of mass kinetic energy.
\beqn
\nonumber
\lefteqn{
\biggl<G_{q}\left(\vec r_{\perp,q}^{\,\prime},z^{\,\prime}\,
|\vec r_{\perp,q},z\right)\,
G_{\bar{q}}^*\left(\vec r_{\perp,\bar q}^{\,\prime},z^{\,\prime}\,
|\vec r_{\perp,\bar q},z\right)\biggr>}
\\
\label{third}& = & \int{\cal D}\mbox{\boldmath$\tau$}_{cm}
\int{\cal D}\mbox{\boldmath$\tau$}_{rel}
\exp\Biggl\{
i\int\limits^{z^{\,\prime}}_{z}d\xi\left(
\frac{\mu}{2}\dot{\mbox{\boldmath$\tau$}}_{rel}^{\,2}
+\frac{\nu}{2}\dot{\mbox{\boldmath$\tau$}}_{cm}^{\,2}\right.\Biggr.\\
&&\nonumber
\qquad\qquad\qquad\qquad\qquad\qquad
+\Biggl.\left.\frac{i}{2}{n_A}\left(b,\xi\right)\sigma_{q\bar{q}}^N
\left(\tau_{rel}\right)
\right)\Biggr\}\\
\label{last}\nonumber & = & 
\frac{\nu}{2\pi i\left(z^{\,\prime}-z\right)}\\\nonumber&\times &
\,\exp\!\!\left(i\frac{\nu}{2}
\frac{\left(
\lambda\left(\vec r_{\perp,q}^{\,\prime}-\vec r_{\perp,q}\right)
+\left(1-\lambda\right)
\left(
\vec r_{\perp,\bar q}^{\,\prime}-\vec r_{\perp,\bar q}
\right)
\right)^2}{z^{\,\prime}-z}\right)\\
&\times &\nonumber
\int\!\!{\cal D}\mbox{\boldmath$\tau$}_{rel}\,
\exp\!\left\{\!
i\int\limits^{z^{\,\prime}}_{z}d\xi\left(
\frac{\mu}{2}\dot{\mbox{\boldmath$\tau$}}_{rel}^{\,2}
+\frac{i}{2}{n_A}\left(b,\xi\right)\sigma_{q\bar{q}}^N
\left(\tau_{rel}\right)
\right)\!\!\right\}.\\
\eeqn
We have introduced the
reduced mass
\beq
\frac{1}{\mu}=\frac{1}{p_{q}^0}+\frac{1}{p_{\bar{q}}^0}
=\frac{1}{\nu\lambda\left(1-\lambda\right)}.
\eeq
Since the imaginary potential depends
only on the relative coordinate, the center of mass propagates freely and what
remains is the effective propagator
\beqn\label{prop}
\lefteqn{W\left(\mbox{\boldmath$\rho$}^{\,\prime},z^{\,\prime}\,
|\,\mbox{\boldmath$\rho$},z\right)}\\&&
=\nonumber\int{\cal D}\mbox{\boldmath$\tau$}_{rel}\,
\exp\left\{
i\int\limits^{z^{\,\prime}}_{z}d\xi\left(
\frac{\mu}{2}\dot{\mbox{\boldmath$\tau$}}_{rel}^{\,2}
-V_{opt}\left(b,\tau_{rel},\xi\right)\right)\right\},
\eeqn
with $\mbox{\boldmath$\rho$}^{\,\prime}=\vec r_{\perp,\bar q}^{\,\prime}
-\vec r_{\perp,q}^{\,\prime}$ and
$\mbox{\boldmath$\rho$}=\vec r_{\perp,\bar q}-\vec r_{\perp,q}$ and the optical potential
\beq
V_{opt}\left(b,\rho,z\right)=
-\frac{i}{2}{n_A}\left(b,z\right)\sigma_{q\bar{q}}^N
\left(\rho\right).
\eeq
It fullfills the equation
\beqn\nonumber
\left[i\frac{\partial}{\partial z^{\,\prime}}
+\frac{\Delta_\perp\left(\rho^{\,\prime}\right)}
{2\nu\lambda\left(1-\lambda\right)}
-V_{opt}\left(b,\rho^{\,\prime},z^{\,\prime}\right)\right]
W\left(\mbox{\boldmath$\rho$}^{\,\prime},z^{\,\prime}\,
|\,\mbox{\boldmath$\rho$},z\right)\\=
i\delta\left(z^{\,\prime}-z\right)
\delta^{\left(2\right)}\left(\mbox{\boldmath$\rho$}^{\,\prime}-\mbox{\boldmath$\rho$}\right).
\qquad\eeqn
The propagator for the center of mass coordinate produces a $\delta$-function
and thus, we obtain from eq.~(\ref{stot}) after averaging over the medium
\beqn\label{stotal}
\nonumber \sigma_{tot}^{\gamma^*A} & = & \frac{Z_f^2\alpha_{em}}{4\nu^2}\,
\int d^2 b\;2\Re\int\limits_{0}^1 
\frac{d\lambda}{\lambda^2\left(1-\lambda\right)^2}
\;\\
\nonumber& & \;\;\times\;\;
\int\limits_{-\infty}^{\infty} dz \int\limits_{z}^{\infty} dz^\prime 
e^{iq_L^{min}\left(z-z^\prime\right)} \\
& & \;\;\times\;\;
{\cal O}\!\left(\mbox{\boldmath$\rho$}\right)
{\cal O}^*\!\!\left(\mbox{\boldmath$\rho$}^{\,\prime}\right)
W\!\!\left(\mbox{\boldmath$\rho$}^{\,\prime},z^{\,\prime}\,
|\,\mbox{\boldmath$\rho$},z\right)
\Big|_{\mbox{\boldmath$\rho$}^{\,\prime}=\mbox{\boldmath$\rho$}=\vec 0},
\eeqn
with
\beq\label{opo}
{\cal O}\left(\mbox{\boldmath$\rho$}\right)
  = 
{\cal O}_T\left(\mbox{\boldmath$\rho$}\right)+
{\cal O}_L\left(\mbox{\boldmath$\rho$}\right),
\eeq
where the transverse part of the operator is
\beqn\nonumber\label{opot}
{\cal O}_T\left(\mbox{\boldmath$\rho$}\right)
  &=& 
\chi_q^\dagger
\displaystyle{\{ }
m_f\mbox{\boldmath$\sigma$}\cdot\mbox{\boldmath$\epsilon$}_T
+i\left(1-2\lambda\right)\mbox{\boldmath$\sigma$}\cdot\vec e_z
\mbox{\boldmath$\epsilon$}_T\cdot\mbox{\boldmath$\nabla$}\left(\mbox{\boldmath$\rho$}\right)
\\&&+
\left(\vec e_z\times\mbox{\boldmath$\epsilon$}_T\right)
\cdot\mbox{\boldmath$\nabla$}\left(\mbox{\boldmath$\rho$}\right)
\}\chi_{\bar q}
\eeqn
and the longitudinal part
\beq\label{opol}
{\cal O}_L\left(\mbox{\boldmath$\rho$}\right)
  = 
\chi_q^\dagger
\displaystyle{\{ }
2\,Q\,\lambda\left(1-\lambda\right)
\}\chi_{\bar q}=2\,Q\,\lambda\left(1-\lambda\right)
\delta_{\lambda_q,-\lambda_{\bar q}}.
\eeq

Equation (\ref{stotal}) is the central result of this work. It is the total
cross section for production of a $q\bar q$-pair from a virtual photon
scattering off a nucleus. We have not summed over the spins of the quark and the
antiquark and not averaged over the polarizations of the photon. We have not
summed over the different flavors, either. 
The expression for the operator ${\cal O}$, eq.~(\ref{opo}),
depends on the spin vector of the
quark and the antiquark. The directions of these vectors may be fixed
arbitrarily.

In order to represent the result in the form (\ref{form}), we rewrite $W$ and
it's complex conjugate in an
expansion. The results can be combined in the following way:
\beqn\nonumber\label{expansion}
\lefteqn{
W\left(\mbox{\boldmath$\rho$}^{\,\prime},z^{\,\prime}\,
|\,\mbox{\boldmath$\rho$},z\right)}\\[1.5ex] \nonumber& = &
W_0\left(\mbox{\boldmath$\rho$}^{\,\prime},z^{\,\prime}\,
|\,\mbox{\boldmath$\rho$},z\right)\\
\nonumber
& + & i\int\limits_{z}^{z^{\,\prime}}dz_1\int d^2\rho_1
W_0\left(\mbox{\boldmath$\rho$}^{\,\prime},z^{\,\prime}\,
|\,\mbox{\boldmath$\rho$}_1,z_1\right)\,\\&&\nonumber\;\;\times\;\;
V_{opt}\left(b,\rho_1,z_1\right)\,
W_0\left(\mbox{\boldmath$\rho$}_1,z_1\,|\,\mbox{\boldmath$\rho$},z\right)\\
\nonumber
& - &\int\limits_{z}^{z^{\,\prime}}dz_1\int d^2\rho_1
\int\limits_{z}^{z_1}dz_2\int d^2\rho_2\\&&\;\;\times\;\;
W_0\left(\mbox{\boldmath$\rho$}^{\,\prime},z^{\,\prime}\,|\,\mbox{\boldmath$\rho$}_1,z_1\right)\,
V_{opt}^*\left(b,\rho_1,z_1\right)\,\\[1.5ex]
& & \;\;\times\;\;\nonumber
W\left(\mbox{\boldmath$\rho$}_1,z_1\,|\,\mbox{\boldmath$\rho$}_2,z_2\right)
V_{opt}\left(b,\rho_2,z_2\right)\,
W_0\left(\mbox{\boldmath$\rho$}_2,z_2\,|\,\mbox{\boldmath$\rho$},z\right).
\eeqn
Here, $W_0$ is the propagator corresponding to (\ref{prop}), when the potential
is absent. The first term gives a divergent contribution, which is the
wave-function renormalization for the photon. The second term leads to the first
contribution in (\ref{form}). The
operators ${\cal O}_{T,L}$ applied to the free propagator $W_0$
give the light-cone wavefunctions
$\Phi_{T,L}\left(\rho,\lambda\right)$, up to a constant overall factor. 
The third term in the above expansion is the
interference term. Since this term contains the full propagator, there are no
higher terms in this expansion. 

As an example for further calculation, consider the integral
\beqn
\nonumber
\cal{I} & = & -
\int\limits_{-\infty}^{\infty} dz \int\limits_{z}^{\infty} 
dz^\prime 
e^{iq_L^{min}\left(z-z^\prime\right)} \\
&&\;\;\times\;\;\nonumber
\int\limits_{z}^{z^{\,\prime}}dz_1\int d^2\rho_1
\int\limits_{z}^{z_1}dz_2\int d^2\rho_2\\
\nonumber &&\;\;\times\;\;
V_{opt}^*\left(b,\rho_1,z_1\right)\,
V_{opt}\left(b,\rho_2,z_2\right)
W_0\left(\mbox{\boldmath$\rho$}^{\,\prime},z^{\,\prime}\,
|\,\mbox{\boldmath$\rho$}_1,z_1\right)\,
\\[1.5ex]
\label{integral}& & \;\;\times\;\;
W\left(\mbox{\boldmath$\rho$}_1,z_1\,|\,\mbox{\boldmath$\rho$}_2,z_2\right)
W_0\left(\mbox{\boldmath$\rho$}_2,z_2\,|\,\mbox{\boldmath$\rho$},z\right),
\eeqn
that is needed to calculate the interference part
\beqn\label{inter}\nonumber
\sigma_{tot}^{int} &=&  \frac{Z_f^2\alpha_{em}}{4\nu^2}\,
\int d^2 b\;2\Re\int\limits_{0}^1 
\frac{d\lambda}{\lambda^2\left(1-\lambda\right)^2}\\&&
{\cal O}\left(\mbox{\boldmath$\rho$}\right)
{\cal O}^*\left(\mbox{\boldmath$\rho$}^{\,\prime}\right)
\;{\cal I}\,
\Big|_{\mbox{\boldmath$\rho$}^{\,\prime}=\mbox{\boldmath$\rho$}= \vec 0}.
\eeqn

With the new variable $\varepsilon^2=\lambda\left(1-\lambda\right)\,Q^2+m_f^2$
we find
\beq
q_L^{min}=\frac{\varepsilon^2}{2\nu\lambda\left(1-\lambda\right)},
\eeq
and because of the relation
\beqn\nonumber{
\left[i\frac{\partial}{\partial z^{\,\prime}}
+\frac{\Delta_\perp\left(\rho^{\,\prime}\right)-\varepsilon^2}
{2\nu\lambda\left(1-\lambda\right)}
\right]
\left(W_0\left(\mbox{\boldmath$\rho$}^{\,\prime},z^{\,\prime}\,|\,\mbox{\boldmath$\rho$},z\right)
e^{-iq_L^{min}\left(z^{\,\prime}-z\right)}\right)}\\=
i\delta\left(z^{\,\prime}-z\right)
\delta^{\left(2\right)}\left(\mbox{\boldmath$\rho$}^{\,\prime}-\mbox{\boldmath$\rho$}\right),
\qquad
\eeqn
we can write the propagator in the form
\beqn\lefteqn{\label{x}
W_0\left(\mbox{\boldmath$\rho$}^{\,\prime},z^{\,\prime}\,|\,\mbox{\boldmath$\rho$},z\right)
e^{-iq_L^{min}\left(z^{\,\prime}-z\right)}}\\&=&\nonumber
\int\frac{d^2 l_\perp}{\left(2\pi\right)^2}
\int\limits_{-\infty}^{\infty}\frac{d \omega}{2\pi}\;
\frac{\exp\left\{-i\omega\left(z^{\,\prime}-z\right)
+i\vec l_\perp\cdot\left(\mbox{\boldmath$\rho$}^{\,\prime}-\mbox{\boldmath$\rho$}\right)
\right\}}{
{\omega-\displaystyle{
\frac{\vec{l}_\perp^2+\varepsilon^2}{2\nu\lambda\left(1-\lambda\right)}}
+i\eta}}.
\eeqn
The $+i\eta$-prescription for the pole in the complex $\omega$-plane ensures
that $W_0\left(\mbox{\boldmath$\rho$}^{\,\prime},z^{\,\prime}\,|\,\mbox{\boldmath$\rho$},z\right)=0$ for
$z>z^{\,\prime}$. Putting (\ref{x}) into (\ref{integral}) yields after a short
calculation
\beqn
\nonumber
\cal{I} & = & 
-\frac{4\nu^2\lambda^2\left(1-\lambda\right)^2}{\left(2\pi\right)^2}\\
\nonumber& & \;\;\times\;\;
\int\limits_{-\infty}^{\infty} dz_1 \int\limits_{-\infty}^{z_1} 
dz_2
e^{iq_L^{min}\left(z_2-z_1\right)} 
\int d^2\rho_1
\int d^2\rho_2\\
\label{integral2}& & \;\;\times\;\;
V_{opt}^*\left(b,\rho_1,z_1\right)\,
V_{opt}\left(b,\rho_2,z_2\right)
\\
\nonumber& & \;\;\times\;\;
K_0\!\left(\varepsilon\left|\mbox{\boldmath$\rho$}^{\,\prime}-\mbox{\boldmath$\rho$}_1\right|\right)
W\left(\mbox{\boldmath$\rho$}_1,z_1\,|\,\mbox{\boldmath$\rho$}_2,z_2\right)
K_0\!\left(\varepsilon\left|\mbox{\boldmath$\rho$}_2-\mbox{\boldmath$\rho$}\right|\right).
\eeqn
$K_0$ is the MacDonald function of zeroth order.
We have used the relation
\beq
K_0\left(\varepsilon\rho\right)=
\frac{1}{2\pi}\int d^2l_\perp\frac{e^{i\vec
l_\perp\cdot\mbox{\boldmath$\rho$}}}{\vec l_\perp^2+\varepsilon^2}.
\eeq
We insert (\ref{integral2}) into (\ref{inter})
and calculate the contribution from the second term in the expansion
(\ref{expansion})
in a similar way. We obtain for the total cross section
\beqn
\nonumber\label{cross}
\sigma_{tot}^{\gamma^*\,A} & = & 
A\,Z_f^2\,\frac{\alpha_{em}}{\left(2\pi\right)^2}
\int\limits_0^1d\lambda\int d^2\rho_1
\,\sigma^N_{q\bar{q}}\left(\rho_1\right)\,\\
\nonumber &\times&
\left|{\cal O}\left(\mbox{\boldmath$\rho$}\right)
K_0\left(\varepsilon\left|\mbox{\boldmath$\rho$}-\mbox{\boldmath$\rho$}_1\right|\right)
\Big|_{\mbox{\boldmath$\rho$}= \vec 0}
\right|^2
\\
\nonumber
& - & Z_f^2\,\frac{\alpha_{em}}{\left(2\pi\right)^2}\,
2\Re\int d^2b
\int\limits_{-\infty}^{\infty} dz_1 \int\limits_{z_1}^{\infty} dz_2\\
\nonumber& & \;\;\times \;\;
\int\limits_0^1d\lambda\int d^2\rho_1\int d^2\rho_2\,
e^{-iq_L^{min}\left(z_2-z_1\right)} \\
& & \;\;\times \;\;\nonumber
V_{opt}^*\left(b,\rho_1,z_1\right)\,
V_{opt}\left(b,\rho_2,z_2\right)
\\
\nonumber& & \;\;\times \;\;
\left({\cal O}^*\left(\mbox{\boldmath$\rho$}^{\,\prime}\right)
K_0\left(\varepsilon\left|\mbox{\boldmath$\rho$}^{\,\prime}-\mbox{\boldmath$\rho$}_1\right|\right)
\Big|_{\mbox{\boldmath$\rho$}^{\,\prime}= \vec 0}\right)\\
\nonumber& & \;\;\times \;\;
W\left(\mbox{\boldmath$\rho$}_2,z_2\,|\,\mbox{\boldmath$\rho$}_1,z_1\right)\,\\
& & \;\;\times \;\;
\left({\cal O}\left(\mbox{\boldmath$\rho$}\right)
K_0\left(\varepsilon\left|\mbox{\boldmath$\rho$}-\mbox{\boldmath$\rho$}_2\right|\right)
\Big|_{\mbox{\boldmath$\rho$}= \vec 0}\right).
\eeqn 

With help of
the relation
\beq
\mbox{\boldmath$\nabla$}\left(\mbox{\boldmath$\rho$}\right)K_0\left(\varepsilon\rho\right)=
-\varepsilon\frac{\mbox{\boldmath$\rho$}}{\rho}K_1\left(\varepsilon\rho\right)
\eeq
we find the light cone wave functions
\beqn
\Phi_T\left(\varepsilon\mbox{\boldmath$\rho$}\right) & = & 
Z_f\,\frac{\sqrt{\alpha_{em}}}{2\pi}
{\cal O}_T\left(\mbox{\boldmath$\rho$}\right)
K_0\left(\varepsilon\rho\right)\\
\nonumber& = &
Z_f\,\frac{\sqrt{\alpha_{em}}}{2\pi}
\Bigg\{m\,K_0\left(\varepsilon\rho\right)\,
\delta_{\lambda_q,\lambda_{\bar q}}
\,\delta_{\lambda_q,\lambda_{\gamma}}\\
& &\nonumber
+\Bigl(i\lambda_q\left(2\lambda-1\right)\mbox{\boldmath$\epsilon$}_T\cdot\vec e_\rho
+\left(\mbox{\boldmath$\epsilon$}_T\times\vec e_z\right)\cdot\vec e_\rho
\Bigr)\\
& &\;\;\times \;\;\varepsilon K_1\left(\varepsilon\rho\right)
\delta_{\lambda_q,-\lambda_{\bar q}}\Bigg\}
\eeqn
and
\beqn
\Phi_L\left(\varepsilon\mbox{\boldmath$\rho$}\right) & = & 
Z_f\,\frac{\sqrt{\alpha_{em}}}{2\pi}
{\cal O}_L\left(\mbox{\boldmath$\rho$}\right)
K_0\left(\varepsilon\rho\right)\\
& = &
Z_f\,\frac{\sqrt{\alpha_{em}}}{2\pi}
2\,Q\,\lambda\left(1-\lambda\right)
\,K_0\left(\varepsilon\rho\right)\,
\delta_{\lambda_q,-\lambda_{\bar q}}
\eeqn
We made use of the Kronecker-$\delta$. 
The unit vector in $\mbox{\boldmath$\rho$}$-direction is denoted by $\vec e_\rho$.
As spin vector we have chosen the unit
vector in $z$-direction and the parameter $\lambda_q$ takes the value $+1$ for
spin in positive $z$-direction and the value $-1$ otherwise. For the antiquark
it is vice versa. For the photon, $\lambda_\gamma=+1$ for positive helicity and
$\lambda_\gamma=-1$ for negative helicity. The transverse light cone wave
function has one part that depends on $K_0$ and another part dependend on $K_1$,
the MacDonald function of first order. Note the different spin structures of
these parts. In the $K_0$-part, the spins of the quarks add up to the spin of
the photon, but in the $K_1$-part of the transverse light-cone wavefunction, 
the spins of the quarks add to $0$ and the
pair gets a spatial angular momentum.
We finally sum over all flavors, colors, helicities and spin states and get 
the following expression:
\beqn
\nonumber\label{final}
\bar\sigma_{tot}^{\gamma^*\,A} & = & A\int\limits_0^1d\lambda\int d^2\rho
\,\sigma^N_{q\bar{q}}\left(\rho\right)\,
\left(\left|\bar\Phi_{T}\left(\varepsilon\rho\right)\right|^2
+\left|\bar\Phi_{L}\left(\varepsilon\rho\right)\right|^2\right)\\
\nonumber
& - & \frac{3\alpha_{em}}{\left(2\pi\right)^2}
\sum\limits_{f=1}^{N_f}Z_f^2\,\Re\int d^2b
\int\limits_{-\infty}^{\infty} dz_1 \int\limits_{z_1}^{\infty} dz_2\\
\nonumber& & \;\;\times \;\;
\int\limits_0^1d\lambda\int d^2\rho_1\int d^2\rho_2\,
e^{-iq_L^{min}\left(z_2-z_1\right)} \\
\nonumber& & \;\;\times \;\;
{n_A}\left(b,z_1\right)\,{n_A}\left(b,z_2\right)\,
\sigma^N_{q\bar{q}}\left(\rho_2\right)\,
\sigma^N_{q\bar{q}}\left(\rho_1\right)
\\
\nonumber& & \;\;\times \;\;
\Biggl\{\left(1-2\lambda\left(1-\lambda\right)\right)\varepsilon^2
\frac{\mbox{\boldmath$\rho$}_1\cdot\mbox{\boldmath$\rho$}_2}{\rho_1\rho_2}
K_1\left(\varepsilon\rho_1\right)K_1\left(\varepsilon\rho_2\right)\Biggr.\\
&&\nonumber\Biggl.
\!\!\!\!\qquad+\left(m_f^2+4Q^2\;\lambda^2\left(1-\lambda\right)^2\right)
K_0\left(\varepsilon\rho_1\right)K_0\left(\varepsilon\rho_2\right)
\Biggr\}\\
& & \;\;\times \;\;
W\left(\mbox{\boldmath$\rho$}_2,z_2\,|\,\mbox{\boldmath$\rho$}_1,z_1\right).
\eeqn 
Here, $\left|\bar\Phi_{T,L}\left(\varepsilon\rho\right)\right|^2$ 
are the absolute
squares of the transverse and the longitudinal light-cone wavefunctions,
sum\-med over all flavors, see
eq.~(\ref{LCWF}).
This form  was used in
\cite{first} for a calculation of nuclear shadowing.
Eq.~(\ref{final}) was for the first time
suggested in a paper by Zakharov
\cite{Slava1}.

Let us summarize the assumptions and approximations entering this derivation.
We start from the Dirac equation with an abelian potential.
 We use this
simplification, since we know, that all dependence of this potential will be put
into the dipole cross section.
Because the final result interpolates between $A\sigma_{tot}^{\gamma^*N}$ and
eq.~(\ref{eikonal}), we assume, that our results hold also for the case
of a nonabelian potential. 

Further, no interaction between 
the quark and the antiquark is taken into account and
therefore, the two Dirac equations decouple.
We use the Furry-Sommerfeld-Maue wavefunctions \cite{Furry,SM} 
that are known to be good
approximations to the continous spectrum of the Dirac equation 
for high energies. Then we derive a two
dimensional Schr\"odinger equation for a scalar function that may be regarded as
an effective wavefunction. The $z$-co\-or\-din\-ate
plays the role of time, since the particles move almost with the velocity of
light. This Schr\"odinger equation is solved in terms of it's Green function.
Averaging over all scattering centers in the nucleus yields an optical potential
that is proportional to the total cross section for scattering a $q\bar q$-pair
off a nucleus. We have omitted the real part of the forward 
scattering amplitude.
All dependence on the potential is absorbed into this cross
section. We propose to use the cross section as input for calculations and not
the potential from which it originates. It may be taken from experimental data
and is the nonperturbative input for our formulae.
Analysis of hadronic cross sections \cite{PH} suggests $\sigma_{q\bar
q}^N(\rho)\approx C\rho^2$ with $C$ between $2.5$ and $3$.
Note however, that we do not make any assumptions on the shape of $\sigma_{q\bar
q}^N(\rho)$ in the derivation.

The averaging procedure and the summation of the multiple scattering
series is similar to the one in Glau\-ber theory
\cite{Glauber} and most of the approximations come in at this point.
First, we neglect all correlations between the nucleons. Then, the influence of
the potential is described by a phase shift
function.
We also assume, that we have a short ranged interaction and the pair interacts
only with one nucleon at a given time. 
It has been demonstrated by Mueller \cite{Al}, that in Born approximation the
dominant contribution comes from graphs, where the $q\bar q$-pair interacts with
the different nucleons one after another via two gluon exchange.
Graphs with crossed gluon lines are suppressed.
This observation justifies our summation procedure.
The phase shift for scattering 
a particle in the pair
off a
single nucleon is calculated for 
an average value of the transverse coordinate of the particle. 
This means, the
transverse coordinates 
should not vary too rapidly within a longitudinal distance of the order
of the interaction range.
When we want to obtain an exponential from the averaging procedure, 
the nuclear
mass number $A$ should be large enough.
Further approximations are, that both particles in the pair see the same nuclear
density. We use the value of the 
density in the middle between the quark and the antiquark
for our calculation. We also approximate the motion 
of the center of mass of the pair by a free motion, since the pair is scattered
predominantly in forward direction.

We finally arrive at the result eq.~(\ref{stotal}). This formula allows to
calculate the cross section $\sigma^{\gamma^*A}_{tot}$ 
for arbitrary polarization of
the photon and the pair. However, this equation is not convenient for numerical
calculations and we modify our result, introducing the light-cone 
wavefunctions, eq.~(\ref{cross}). Since we are not interested in certain
polarizations for a calculation of nuclear shadowing, we sum over all helicity
and spin states, arriving at eq.~(\ref{final}).

\section{Summary and conclusions}
\label{sec:3}
 
We have considered nuclear shadowing in the rest frame of the nucleus, in which
the virtual photon fluctuates into a $q\bar q$-pair. 
In the preceding section we gave a detailed derivation of a formula for nuclear
shadowing in DIS that accounts for both, the finite lifetime of the hadronic
fluctuation and all multiple scattering terms. With this formula,
eq.~(\ref{final}), it is possible
to calculate nuclear shadowing for
moderate values of $x_B$, 
$x_B > 0.01$, where the lifetime of the fluctuation does not
exceed the nuclear radius by orders of magnitude.
It must be clearly emphasized, that
we have only taken the $q\bar q$-Fock component of the
photon into account and thus, the applicability of our results
is restricted to values of $x_B$ and $Q^2$, where corrections from higher 
Fock-states of the photon, containing gluons, are not important.
In particular,
the structure function $F_2$ of the proton, calculated
in our model, does not depend on $x_B$. 
In order to get the steep rise at very small $x_B$, one has to take higher
Fock-states of the photon into account.
Further, shadowing for the longitudinal cross section drops as $1/Q^2$ as
$Q^2\rightarrow\infty$, since we have no gluon shadowing in our model.
Adding one gluon to the $q\bar q$-pair would change this.
This problem will be addressed in a forthcoming paper.

All assumptions summarized in the end of sec.~\ref{sec:2}
seem reasonable to us and the approximations should 
work as long as the nuclear mass number $A$ is not too small.

Numerical calculations \cite{first} with eq.~(\ref{final}) show that higher
order scattering terms have a significant influence on the total cross section,
especially for heavy nuclei. In \cite{first}, the formula for nuclear shadowing 
was given without derivation. 
This has now been made up.

The cross section $\sigma^{\gamma^*A}_{tot}$ is identical to the total cross
section for production of a $q\bar q$-pair from the virtual photon in the field
of the nucleus. The suppression occurs because of destructive interference
between pairs created and different longitudinal coordinates within the
coherence length. 
Thus, shadowing may be regarded as the
Landau-Pomeranchuk-Migdal-effect \cite{Landau,Migdal}  for pair production. 
This effect is the analog of the more widely 
known effect for bremsstrahlung and was
first mentioned by Mig\-dal \cite{Migdal} for electron-positron pair production
in  condensed matter. However, this effect 
is practically not observable, because of the low
density of solids. Since the density of nuclear matter is much higher, the
effect occurs in $q\bar q$ pair production in DIS.
For this reason,
eq.~(\ref{final}) was discovered independently by Zakharov
\cite{Slava1}, who considered the LPM-effect for finite size targets.

\begin{acknowledgement} 
We are grateful for stimulating
discussions to J\"org
H\"ufner and Boris Kopeliovich who read the paper and made many useful
comments. 
We thank the MPI f\"ur Kernphysik, Heidelberg, for hospitality.

The work of J.R. and A.V.T. was supported by the Ge\-sell\-schaft f\"ur 
Schwer\-ionen\-for\-schung, GSI, grant HD H\"UF T.
\end{acknowledgement}

\end{document}